\documentclass[12pt]{article}
\usepackage{graphicx}
\usepackage[euler]{textgreek}
\usepackage{cite}

\newcommand\pubnumber{NuPhys2018-Valder}
\newcommand\pubdate{\today}

\def\napoli{Department of Physics\\
University of Warwick, Coventry CV4 7AL, United Kingdom}
\def\support{\footnote{On behalf of the Hyperk-Kamiokande proto-collaboration}\textsuperscript{,}\footnote{Supported by the UK Science and Technologies Faculties Council}}

\def\Title#1{\begin{center} {\Large #1 } \end{center}}
\def\Author#1{\begin{center}{ \sc #1} \end{center}}
\def\Address#1{\begin{center}{ \it #1} \end{center}}

\newcommand\pubblock{\rightline{\begin{tabular}{l} \pubnumber\\
         \pubdate  \end{tabular}}}
\newenvironment{Abstract}{\begin{quotation}  }{\end{quotation}}
\newenvironment{Presented}{\begin{quotation} \begin{center} 
             PRESENTED AT\end{center}\bigskip 
      \begin{center}\begin{large}}{\end{large}\end{center} \end{quotation}}
\def\Acknowledgements{\bigskip  \bigskip \begin{center} \begin{large}
             \bf ACKNOWLEDGEMENTS \end{large}\end{center}}

\def\beq{\begin{equation}}
\def\eeq#1{\label{#1}\end{equation}}
\def\eeqn{\end{equation}}

\def\beqa{\begin{eqnarray}}
\def\eeqa#1{\label{#1}\end{eqnarray}}
\def\eeqan{\end{eqnarray}}

\let\bar=\overbar

\def\Dslash{\not{\hbox{\kern-4pt $D$}}}
\def\dslash{\not{\hbox{\kern-2pt $\del$}}}

\def\msb{{\bar{\ssstyle M \kern -1pt S}}}

\begin{document}
\begin{titlepage}
\pubblock

\vfill
\Title{Diffuser Research and Development for Optical Calibration Systems in Hyper-Kamiokande}
\vfill
\Author{ Sammy Valder\support}
\Address{\napoli}
\vfill
\begin{Abstract}
Proposed as part of the next generation of water Cherenkov detectors, Hyper-Kamiokande will have a vastly improved potential in determining leptonic CP violation in neutrino oscillations. Well understood optical diffusers are needed as part of an integrated light injection system for calibration of ultrasenstive photo-detectors. Research and development into optical diffuser technology is presented and summaries of recent installations in Super-Kamiokande are discussed. 
\end{Abstract}
\vfill
\begin{Presented}
NuPhys 2018, Prospects in Neutrino Physics\\
Cavendish Conference Centre, London, UK, December 19--21, 2018
\end{Presented}
\vfill
\end{titlepage}
\def\thefootnote{\fnsymbol{footnote}}
\setcounter{footnote}{0}

\section{Introduction}

A new generation of water Cherenkov neutrino detectors is fast approaching. With construction set to begin in 2020, Hyper-Kamiokande (Hyper-K) will have a vastly improved potential in determining leptonic CP violation in neutrino oscillations \cite{abe2011letter}. At 187kT fiducial mass, Hyper-Kamiokande is set to be the largest world's water Chekerenov detector, Eight times larger than Super-Kamiokande (Super-K), the largest water Cherenkov detector currently in operation \cite{abe2011letter}. Surrounding Hyper-Kamiokande's 60m x 74m cylindrical tank will be in order of 40000 high sensitivity photo detectors, which all need optical calibration. The Hyper-Kamiokande physics goals dictate that we understand the detector to the level of a few percent which can only be achieved with careful calibration systems. Optical diffusers provide an ideal way to calibrate timing and energy information for photo-sensors in large scale detector projects such as Hyper-K. In this report, ongoing work into the research and development of optical diffusers will be discussed. Furthermore recent deployments of such optical calibration systems in Super-K will be outlined.

\section{Calibration Optics}

The light injection system developed for Hyper-K uses an LED/laser source to guide photons through a 200\textmu m core graded index optical fibre cable into the detector. From here there are three main optical components used to manipulate the light for calibration measurements. The first uses the bare end of a fibre as a control light source. The second consists of a collimator, providing a narrow beam designed to illuminate of order 5 inner-detector photomultiplier tubes (PMT). The collimator's main task is to measure properties of the water such as attenuation and photon scattering lengths. For more information on this topic please refer to B. Vinnings work at this conference. The third and final optical component is the wide beam optical diffuser. The main objectives for the diffuser is to provide a well understood distribution of light over a large range of angles in order to provide inter-PMT and PMT timing calibration measurements. It is this diffuser measurement that is the focus of the work reported here.

\section{Wide Beam Diffuser}

As discussed in section 2, the diffuser is used for inter-PMT and timing calibrations. Both of these measurements require a relatively uniform light source over a large range of angles. This is a non-trivial objective over the large range of angles needed for Hyper-Kamiokande, however it is possible to achieve a well-understood, near uniform, distribution over such a range. To achieve this Polymethyl methacrylate (PMMA) was chosen as the material to make the diffuser. A variety of different materials were tested for uniformity and transparency over a range of wavelengths. The optical diffuser is needed to perform over the visible spectrum as well as in the near visible UV spectrum \cite{fukuda2003super}. Out of all the materials tested, PMMA provided the best results in transmitting light over the wavelengths of 300nm - 700nm. 

\begin{figure}[htb]
\centering
\includegraphics[height=1.3in]{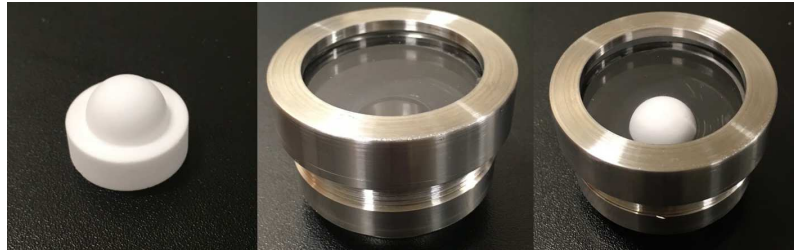}
\caption{The PMMA diffuser outside its enclosure (left). The diffuser enclosure (centre). The diffuser inside its enclosure (right).}
\label{fig:diffuser}
\end{figure}

One disadvantage to using PMMA is its high absorbency and variable light transmission properties when wet \cite{lin1994temperature}. To mitigate this problems the diffuser is housed inside a water-tight stainless steel enclosure, shown in figure \ref{fig:diffuser}. Water-tightness is ensured through a number of rubber O-rings and a coating of epoxy resin glue applied at each potential opening. The enclosure has successfully passed water pressure tests up to pressure of 5 Bar. Further tests at higher pressures are scheduled for the near future. 

As well as the material, a number of shapes for the diffuser ball were considered and tested before construction. The shape of the diffuser needs to provide structural support together with ensuring uniformity in the light distribution. A hemispherical diffuser shape provides both ensuring the possibility for maximum light injection, avoiding a loss due to back-scattering seen in full spherical diffusers. One natural consequence of having a hemispherical shape is the inability to produce a uniform distribution due to inherent geometries. Nevertheless, so long as the light intensity distribution is well-understood and reproducible, this can be accounted for during analysis.

\section{Results} \label{sec::results}

\subsection{Laboratory Measurements} \label{sec::lab}

\begin{figure}[htb]
\centering
\includegraphics[height=1.7in]{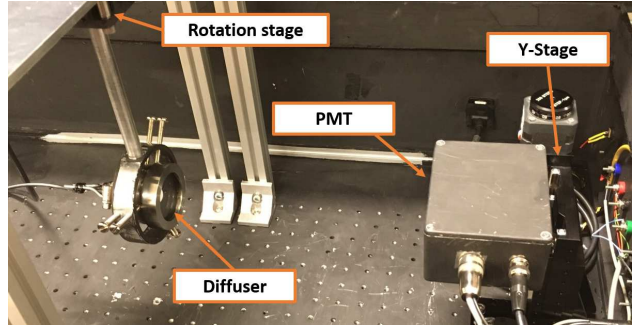}
\caption{The experimental setup used at the University of Warwick Hyper-K laboratory to measure the light intensity and timing distributions of each diffuser.}
\label{fig:lab}
\end{figure}

An experimental testbench developed at the University of Warwick has been designed to test the uniformity of optical diffuser light transmission profiles. Housed inside a dark box to minimise optical reflections, the setup is shown in figure \ref{fig:lab}. The diffuser is held on a rotation stage allowing the full horizontal plane to be scanned over $\pm$90 degrees from the normal. Light injection is provided by individually selectable pulsed semiconductor lasers of wavelengths 450nm and 520nm. A 200\textmu m core step-index fibre optic cable couples the laser with the diffuser. A photomultiplier tube (PMT) is used to measure the light intensity and timing distributions from the diffuser over a range of angles. A second degree of freedom is achieved by mounting the PMT on a Y-Stage. This allows for horizontal scans at different heights relative to the diffuser. 

\begin{figure}[htb]
\centering
\includegraphics[height=2in]{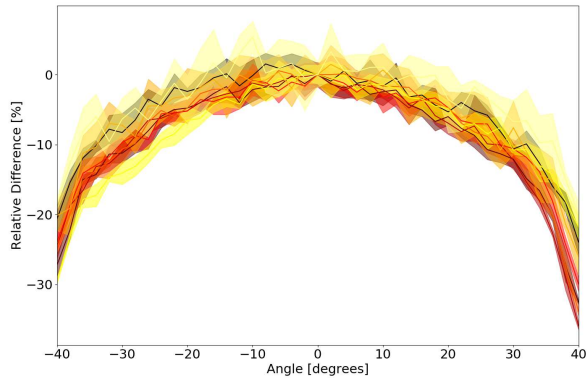}
\caption{The light intensity distributions for all 7 diffusers designed for the Super-K installation. Each distribution is normalised to the zero angle. }
\label{fig:results}
\end{figure}

Angular scans of diffuser output result in measurements of the intensity distribution from each diffuser. Figure \ref{fig:results} shows the results of angle scans for seven different diffusers manufactured for the Super-K installation described in section \ref{sec::sk}. Figure \ref{fig:results} demonstrates that the hemispherical PMMA diffuser ball emits a reproducible and consistent light intensity distribution. This distribution is well understood and can be successfully modelled with hemispherical geometry simulations. The drop off seen above $\pm$30 degrees is due to the inherent field-of-view from the diffuser enclosure.

\subsection{Super-Kamiokande Deployment} \label{sec::sk}

Multiple deployments of the calibration optics have been undertaken inside the Super-Kamiokande detector. The first of which was a test deployment in January 2018. The optics were lowered into the top of the Super-K tank via a mounting plate through the calibration port. Six months of data was taken and validation analyses were performed to confirm and compare the performance of the optics to laboratory measurements seen in section \ref{sec::lab}. This also served as a trial for a long term deployment later in the year.

In the summer 2018, the several optical calibration systems were installed inside the Super-K inner tank. Five identical systems were installed at regular intervals, B1-B5, from the top of the tank, defined as B1, to the bottom of the tank, B5 \cite{abe2014calibration}. Data taking is scheduled to begin in March 2019 and will continue long term. Analysis of the data is also scheduled to begin in April 2019.

\Acknowledgements
We are grateful to the Hyper-K UK calibration collaboration for their combined efforts in the work above. Furthermore we like to give out thanks to the Kamioka Observatory, the Super-Kamiokande experiment and the Super-Kamiokande calibration group for their help in installation and data-taking during 2018.

\bibliographystyle{ap4}

\end{document}